# Effects of CVD Growth Parameters on Global and Local Optical Properties of MoS$_2$ Monolayers

Ana Senkić[1,2*], Josip Bajo[3*], Antonio Supina[1,2], Borna Radatović[1], and Nataša Vujičić[1*]

*[1] Institute of Physics, Center for Advanced Laser Techniques and Center of Excellence for Advanced Materials and Sensing Devices, Zagreb 10 000, Croatia*
*[2] Faculty of Physics, University of Rijeka, Rijeka 51 000, Croatia*
*[3] University of Vienna, Faculty of Physics, Vienna Center for Quantum Science and Technology (VCQ), 1090 Vienna, Austria*

*Address correspondence to Ana Senkić, asenkic@ifs.hr; Nataša Vujičić, natasav@ifs.hr
• These authors contributed equally to this work.

## ABSTRACT

Semiconducting transition metal dichalcogenides (TMDs) combine strong light-matter interaction with good chemical stability and scalable fabrication techniques, and are thus excellent prospects for optoelectronic, photonic and light-harvesting applications. Controllable fabrication of high-quality TMD monolayers with low defect content is still challenging and hinders their adoption for technological application. The optical properties of chemical vapor deposition (CVD) grown monolayer MoS$_2$ are largely influenced by the stoichiometry during CVD by controlled sulfurization of molybdenum (Mo) precursors. Here, we investigate how the sulfur concentration influences the sample morphology and, both globally and locally, their optical response. We confirm that samples grown under a Mo:S > 1:2 stoichiometric ratio have regular morphology facilitated by a moderate coverage of triangular monocrystals with excellent optical response. Our data-driven approach correlates growth conditions with crystal morphology and its optical response, providing a practical and necessary pathway to address the challenges towards the controlled synthesis of 2D TMDs and their alloys with desired optical and electronic properties.

**KEYWORDS** Molybdenum disulfide, Chemical Vapor Deposition (CVD), Growth Parameters, Optical Properties of CVD MoS$_2$

**Introduction**

The diversity of 2D material systems has been increasing substantially over the last two decades, and each new system offers new and exciting research opportunies. Transition metal dichalcogenides (TMDs) form a compelling class of 2D materials of significant interest due to their thickness-dependent electrical and optical properties with potential applications in optoelectronics, flexible electronics, chemical sensing and quantum technologies [1]–[9].

At the monolayer limit, the semiconducting TMDs exhibit direct band gaps within the visible range and large exciton binding energies [10], [11]. The lack of inversion symmetry in the monolayer results in unique optical selection rules, enabling circularly polarized light to selectively populate degenerate valleys in the Brillouin zone, providing an additional degree-of-freedom for the realization of valleytronic devices [3], [12]–[14]. Despite the vast amount of research conducted on TMDs, mostly on $MoS_2$, reproducible fabrication of high-quality TMD monolayers is still challenging. Although the top-down method, such as exfoliation from bulk crystals [15], [16] offers pristine sample quality and proves a satisfactory platform for one-off studies of properties, it is neither scalable nor practical for TMD crystal incorporation into devices. To this end, considerable effort has been taken towards controllable synthesis of monolayer materials using a variety of techniques such as liquid exfoliation method [17]–[19], physical vapor deposition (PVD) method, [20] chemical vapor deposition (CVD) method [21]–[25] and ultra-high vacuum (UHV) molecular beam epitaxy (MBE) [26]–[29] method. As one of the bottom-up techniques, MBE allows the epitaxial growth of 2D van der Waals materials and offers unparalleled control of sample cleanliness and growth parameters due to the UHV environment. However, need for UHV conditions makes it technically more demanding and time consuming. Another limitation of MBE techniques is the need for the growth substrates that facilitates the MBE growth, which can be incompatible with the desired application in optoelectronic devices [30]. The atmospheric pressure CVD is the most widely used technique for the synthesis of TMDs layers because of its simplicity, low costs, low precursor expenditure, fast growth rate and large domain sites making it compatible with industry standards [31]. The CVD allows us an arbitrary choice of a substrate, as well as to control the number of layers and domain size. However, sample uniformity and reproducibility issues have been reported [32], with significant variations not only from the sample to the sample but also across the single grain, posing a significant barrier for the implementation of CVD grown TMDs in technological applications. Variations in optical response and charge transport mechanisms in the CVD grown samples have been attributed to different origins, such as trapped charges at sulfur vacancies [33], grain boundaries [34], trapped charges at the interface of $MoS_2$ and oxide dielectrics [35], extrinsic disorder from adsorbates [36], and other defects within the films. This calls for mensurable and controllable fabrication of TMD monolayers with high-quality optical response and low defect density.

The scope of this work was to investigate how the sulfur concentration in correlation with Mo-precursor concentration influences the morphology and optical response of $MoS_2$ monolayers. The CVD grown samples from liquid-based molybdenum (Mo) precursors were synthesized under different growth conditions and their overall optical properties were investigated immediately upon synthesis by measuring photoluminescence (PL) and Raman signal on different flakes across the sample and from different sample zones. Described workflow includes sample synthesis under certain growth conditions followed by optical spectra acquisition and subsequent data analysis providing a crucial link to determine the next optimal CVD growth set of parameters to produce a high-quality sample with desired morphology [37]. Such an approach, referred as "active learning", allows us to improve the knowledge about the future growth parameters space, its resulting morphology accompanied by sample properties with as few experimental runs as possible [38].

Here, the growth conditions are determinate by three parameters: growth temperature ($T_G$), sulfur temperature ($T_S$) and inert gas flow ($\zeta$). We noted how the optical properties of CVD grown monolayer $MoS_2$ are largely influenced by stoichiometry via controlled sulfurization of Mo-precursors and we correlated the stoichiometry with the optical properties of CVD synthesized monolayer $MoS_2$ flakes prepared under varying degrees of $MoO_3$ sulfurization. When the Mo:S ratio changes and the amount of molybdenum increases, triangular flakes with Mo-zz termination edge form. On the contrary, when the sulfur ratio increases, triangular flakes with only S-zz termination edge form [39]. Therefore, controlled sulfurization of the Mo- precursors leads toward high-quality single crystalline growth with no evidence of grain boundaries or extrinsic disorder from the adsorbates, which have been confirmed with AFM and SEM measurements.

## 1. Materials and methods

### 1.1 $MoS_2$ Synthesis

Figure 1 shows an illustration of the chemical vapor deposition (CVD) setup for $MoS_2$ growth along with a photograph of the sample depicting the droplet deposition contour in the wafer center and the main steps of the synthesis at different stages. The wafer of 10 mm × 10 mm in size is tentatively divided into 4 sample zones marked by dashed lines, with the first zone being upstream, i.e., in the direction of the inert gas flow, as indicated with the arrow, see Fig. 1 b). The first zone includes the region from the nearest wafer edge (with the respect to sulfur boat), over the droplet edge and its depletion zone toward the end of the first quarter of the sample. The second zone continues from the end of the zone 1 toward the droplet center; the third zone continues from the droplet center up to the third quarter of the substrate and the fourth zone continues from the end of the third zone over the outermost droplet edge, its depletion zone until the end the wafer, therefore last quarter of the substrate. The four-zone approach enables us to examine how the sulfur inflow influences crystal growth on different parts of the wafer, having in mind that the sulfur precursor concentration is regulated by wafer distance from the sulfur boat (i.e. from the source of the sulfur), its evaporation temperature and inert gas flow, under the assumption that Mo- precursor and growth promoter are evenly deposited on the substrate over the all four zones. Since evaporation temperature and inert gas flow can be precisely regulated, in order to ensure a systematic growth method with high reproducibility over the whole wafer, with desired crystal morphology and optical properties, the quality of the crystal was checked on a randomly chosen flake from each of the four zones.

Prior to the synthesis, the $Si/SiO_2$ substrate was a blow cleaned against dust particles using argon gas. The silicon part of the substrate was heated with a propane flame in order to minimize any organic impurities. As a Mo- precursor, we used a mixture of two deionized (DI) water-based solutions in equal parts: ammonium heptamolybdate $((NH_4)_6Mo_7O_{24})$ (AHM) solution (Kemika), 15.4 parts per million (ppm) and sodium-molybdate $(Na_2MoO_4)$ solution (Merk's Reagenzien), 15.4 ppm concentration. Our preliminary work showed that a mixture of these two solutions in the same volume parts gives the best crystal morphology and good optical response [40]. When using only AHM solution as the Mo- precursor, grown $MoS_2$ flakes were relatively small and irregular but having fair PL emission spectra. The density of crystal growth along the four substrate zones was moderate and uniform. In the case when only sodium-molybdate solution was used, grown $MoS_2$ islands formed polycrystalline clusters and emission PL spectra were poor. The crystal growth along the substrate zones was excessively dense, creating the inevitable bulky film especially on the droplet edge, confirming that such highly efficient growth is facilitated by sodium catalysts [41].

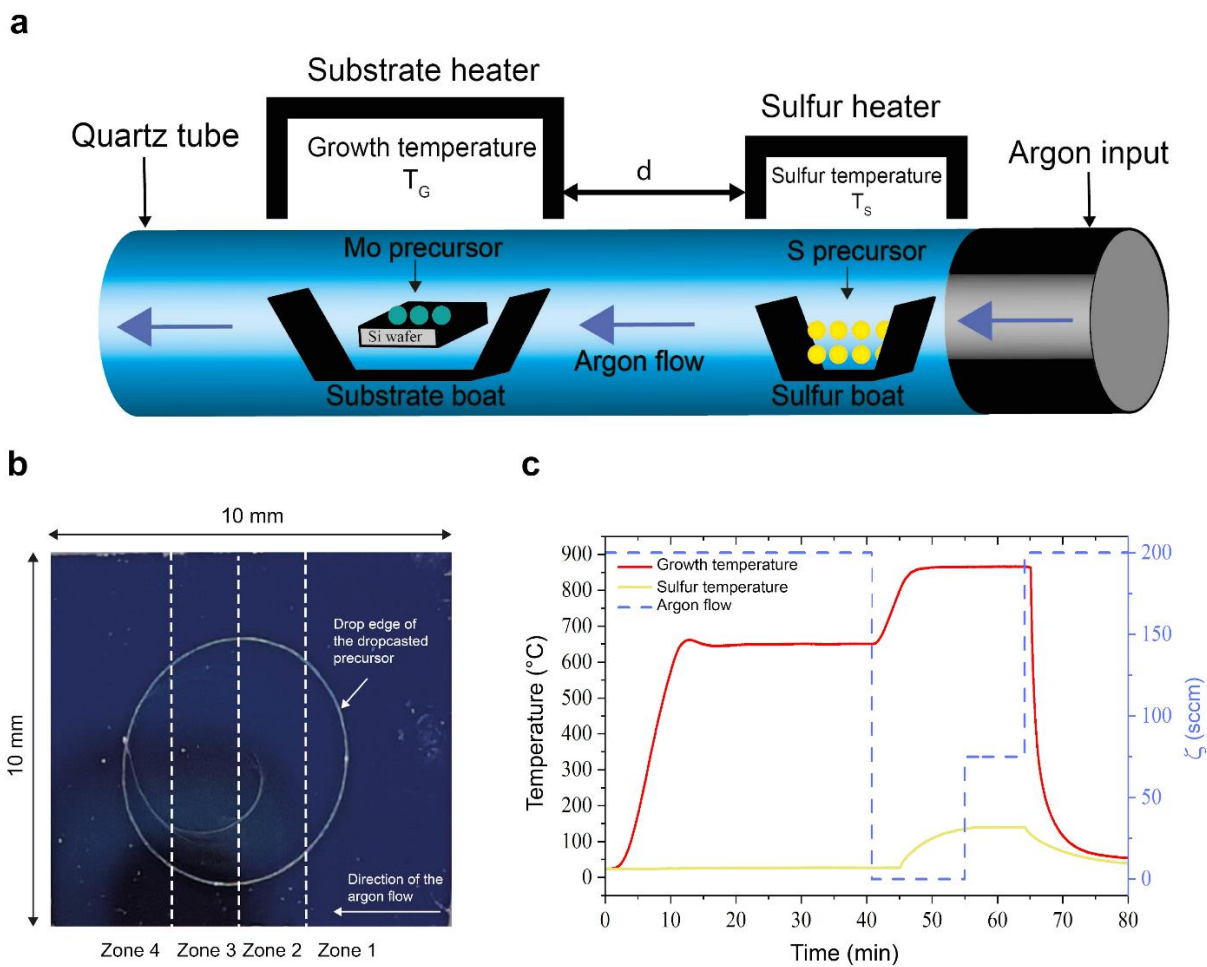

**Fig. 1** Schematic illustration of the CVD growth method based on solution-based metal precursors. (a) Photograph of the sample depicting the droplet deposition contour in the wafer center. The direction of the gas flow during deposition is indicated with an arrow (from the right to the left). Dashed lines indicate zone borders that separate four different zones (1-4) on the wafer, with the first zone being upstream. (b) Profiles of the temperature and argon flow rate programmed for the formation of $MoS_2$ flakes (c)

The proposed chemical reaction for obtaining the $MoO_3$ film as an intermediate Mo- precursor from two water-based salt solutions is given by (Eq. 1) [42]:

$$(NH_4)_6Mo_7O_{24} \times 4H_2O + Na_2MoO_4 \times 2H_2O \rightarrow 8MoO_3 + 8H_2O + 2NaOH + 6NH_3. \qquad (Eq. 1)$$

Residual NaOH and $NH_3$ cause the $SiO_2$ substrate etching, which effects will be discussed later. The metal precursor supply strategy that facilitates the use of these two solution mixtures in the same volume parts results in the moderate growth that characterize large scale uniformly distribution of $MoS_2$ flakes with a lateral size of approximately 50 µm. The obtained crystals have enhanced PL intensity compared to samples synthesized from separate salt solutions. A 10µL droplet of this mixture was dropcast on the cleaned substrate and then placed on a hot plate at 120°C until the droplet has dried (less than one minute). The substrate with the deposited metal precursor was then placed in the CVD furnace at 650°C temperature for 30 minutes under argon flow (see Fig. 1 (c)). The sulfur powder (Sigma-Aldrich) was held in a sulfur boat outside the furnace and a separate heater was used to control the sulfur temperature. After the first phase, the argon flow rate ($\zeta$) was stopped ($\zeta$= 0 sccm) and the furnace temperature was raised to the growth temperature ($T_G$). Also, the sulfur heater was turned on and set to the sulfur evaporation temperature ($T_S$). When both temperatures reached their targeted value, the argon flow rate was set to desired value - between 50 and 100 sccm, and the synthesis process starts. The overall time of synthesis was 10 minutes. When the synthesis was completed, the furnace was cooled down to the constant temperature of 650°C. To achieve rapid cooling, the sample was removed outside of the heater zone, while still being kept inside of the quartz tube. When the sample temperature reached ∼ 200°C, the sample was removed from the CVD furnace.

The presence of seeding materials can favor the formation of initial nucleation sites that will foster the lateral growth of $MoS_2$ crystals, enabling significant improvements in the surface coverage and crystallinity. Rare (dense) nucleation centers will result in poor (high) sample coverage, see Fig. S-1 in the Electronic Supplementary Material (ESM). The CVD parameters can be tuned to obtain a sufficient number of nuclei that will ultimately grow in single-crystal structures with moderate sample coverage. It was demonstrated [43] that the ratio of Mo and S affects the resultant shape and edge structure of $MoS_2$ crystals. When the ratio changes and the amount of molybdenum increases, triangular flakes with Mo-zz termination edge form. Conversely, when the sulfur ratio increases, triangular flakes with S-zz termination edge form. Ye *et al.* [44] analyzed both the in-plane growth of a layer and the formation of vertically stacked layers in relation to the growth temperature and precursor mass flow rate. Their results suggest the increase of the growth temperature and decrease of the inert gas flow rate foster the growth of a newly growing layer (the epitaxial grown second layer). Therefore, it was shown that by tuning the synthesis parameters we can design an effective growth process [45].

In this work, we tuned all three synthesis parameters:

• growth temperature, $T_G$, from 775°C to 925°C;

• sulfur temperature, $T_S$, from 135°C to 145°C;

• argon flow rate, $\zeta$, from 50 sccm to 100 sccm

which yields in synthesis of 30 different samples. Table S-2 in the ESM contains growth parameters for the mentioned 30 samples sorted into 5 batches. By increasing the $T_G$, the amount of molybdenum increases due to the formation of initial nucleation sites; by increasing the $T_S$ overall sulfur concentration increases while the increase of argon flow rate $\zeta$ (i.e. sulfur mass flow rate) causes higher sulfurization and promotes lateral sample growth. Well-balanced set of growth parameters results in moderate sample coverage with the crystals of desired shape and size and high-quality optical properties. To examine the optical response of the synthesized crystals over the entire substrate, we conducted optical measurements on each randomly chosen flake – one at each zone for every sample that was synthesized in order to obtain good statistics. The lateral size of the chosen flakes was up to 50 μm.

## 1.2 Optical Characterization

Optical characterization was performed using a confocal microscope in a backscattered configuration, which serves for photoluminescence (PL) and Raman spectra measurements, as well as for optical mapping of samples. The excitation laser energy was 2.33 eV (532 nm) and the laser power on the sample was 500 μW.

The incident laser was focused by 50× infinity corrected objective (NA = 0.75). The backscattered optical signal was guided through a cage system consisting of a set of Bragg filters and it is coupled in 50 μm core fiber that serves as a confocal detection pinhole. Backscattered light is analyzed in a 50 cm long spectrometer (Andor-500i-B1) equipped with three different diffraction gratings (150, 300, 1800 l mm$^{-1}$) and cooled EM CCD detector (Newton 971). All measurements were conducted at the room temperature. Optical micrographs (OMs) were taken with a commercial optical microscope Leica DM2700M, Leica Microsystems, Wetzlar, Germany.

## 1.3. Scanning Electron Microscope (SEM)

Scanning electron microscope (SEM) Tescan VEGA3 with a tungsten cathode was used for observation of $MoS_2$. Imaging was performed at a working distance of 10 mm and with 5 kV accelerating voltage in resolution mode, while the image acquisition was done with a secondary electron (SE) detector.

## 1.4. Atomic force microscope (AFM)

Atomic force microscope (AFM) images of $MoS_2$ were taken with a JPK Nanowizard Ultra Speed AFM under ambient conditions. Non-contact AC mode was used for the acquisition with a set point of around 70 %. NCHPt tips from Nanoworld with a nominal spring constant of 42 N/m were used. Images were processed with JPK Data Processing software.

## 2. Results and Discussion

### 2.1 Morphology Evolution

Figure 2 shows the morphological evolution of $MoS_2$ flakes growth for different growth temperature $T_G$ and different sulfur evaporation temperature $T_S$ for fixed carrier gas flow rate of 50 sccm. $T_G$ increases in the range from 800°C up to 900°C, in steps of 25°C, while $T_S$ is changed in the range from 135°C up to 145°C, with 5°C increment. The arrows in Fig. 2 indicate the direction along which temperatures $T_G$ and $T_S$ increase. For lower values of $T_G$, $MoS_2$ monolayers evolve to dendritic morphology as the $T_S$ increases, see Fig.2 (a, f, k) and Fig. 2 (b, g, l) indicating S-rich growth conditions. Triangular morphology is restored as $T_G$ increases, see Fig. 2 (i, j) and Fig. 2 (n, o).

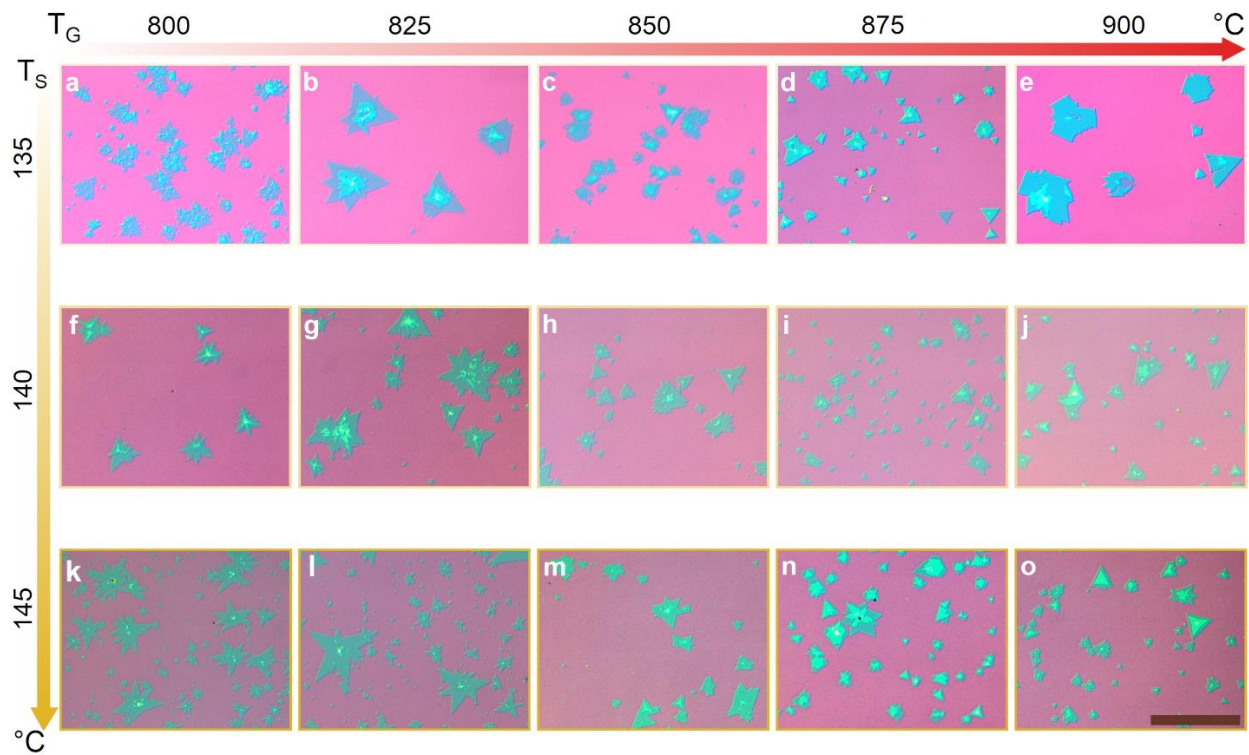

**Fig. 2** OM images of as-grown $MoS_2$ samples as a function of growth temperature ($T_G$) and set sulfur evaporation temperature ($T_S$). The carrier gas flow rate was 50 sccm. Scale bar is 100 μm

An increase of the $T_G$ increases the amount of $MoO_3/MoO_x$ within the region of synthesis [46] that, in turn, gives rise to equilateral triangular $MoS_2$ flake growth for a given degree of sulfurization. Simultaneously, lower number of nucleation centers promotes the growth of the isolated flakes rather than the growth of the partially or continuously merged flakes [41]. For example, since the Mo-precursor concentration is too high at the droplet edges, we were only able to achieve isolated monolayer growth upon high growth temperatures. Otherwise, bulky growth at droplet edges with an inherent depletion zone of about ten micrometers is observed. Further increase of $T_G$ (875-900 °C) leads to more pronounced bilayer and multilayer growth, since the evaporation of Mo- precursors reduces the probability for monolayer growth, promoting the epitaxial growth of additional layer(s) on the top of the first layer, activating the same nucleation centers. [31][41][46][47] Our preliminary work [40] showed that higher variation of $T_S$ up to 170°C leads toward three-point star shape $MoS_2$ flakes and dense growth, see Fig. S-3 in the ESM. This type of flake morphology is typical for S-rich atmosphere and low $T_G$, which can be also tuned toward triangular morphology by decreasing $T_S$ or by increasing $T_G$. Regarding morphology evolution, the best results were obtained for $T_S$ span ranging between 135°C up to 145°C, therefore for the moderate sulfurization during growth process. Multilayer growth is obtained mostly at high growth temperatures. However, if the sulfur temperature is low, multilayer $MoS_2$ islands are obtained throughout the growth temperature range, as demonstrated in [41].

Secondly, we investigated how the growth temperature $T_G$ and different argon flow rates $\zeta$ influence the sample morphology and quality, while maintaining fixed $T_S = 140°C$, as shown in Fig. 3. $T_G$ increases in the range from 800°C up to 900°C, in steps of 25°C, while $\zeta$ is changed in the range from 50 sccm up to 100 sccm with 25 sccm increment. The arrows in Fig. 3 show the direction along which temperature $T_G$ and argon flow $\zeta$ increase. For lower values of $T_G$, increase of the argon flow rate results in a lateral size increase of the flake and more nucleation sites are activated, see Fig. 3 (a, f, k) and Fig. 3 (b, g, l). There is a limit to the value of the argon flow rate, above which no significant increase of the lateral growth is observed, as can be seen in Figs. 3 (k, l). For such a high carrier gas flow rate, the $MoS_2$ crystals are more likely to grow under kinetic conditions rather than under thermodynamic ones [43] resulting in irregular, dendritic flakes, especially if the bilayer growth temperature is not high enough. The high flow rate promotes the mass transfer process, which contributes to the increase in the crystal growth rate, see Fig. 3 (a, f, k), Fig. 3 (b, g, l) and Fig. 3 (c, h, m) with the formation of dendritic morphologies, which are unfavorable in producing high-quality 2D crystals.

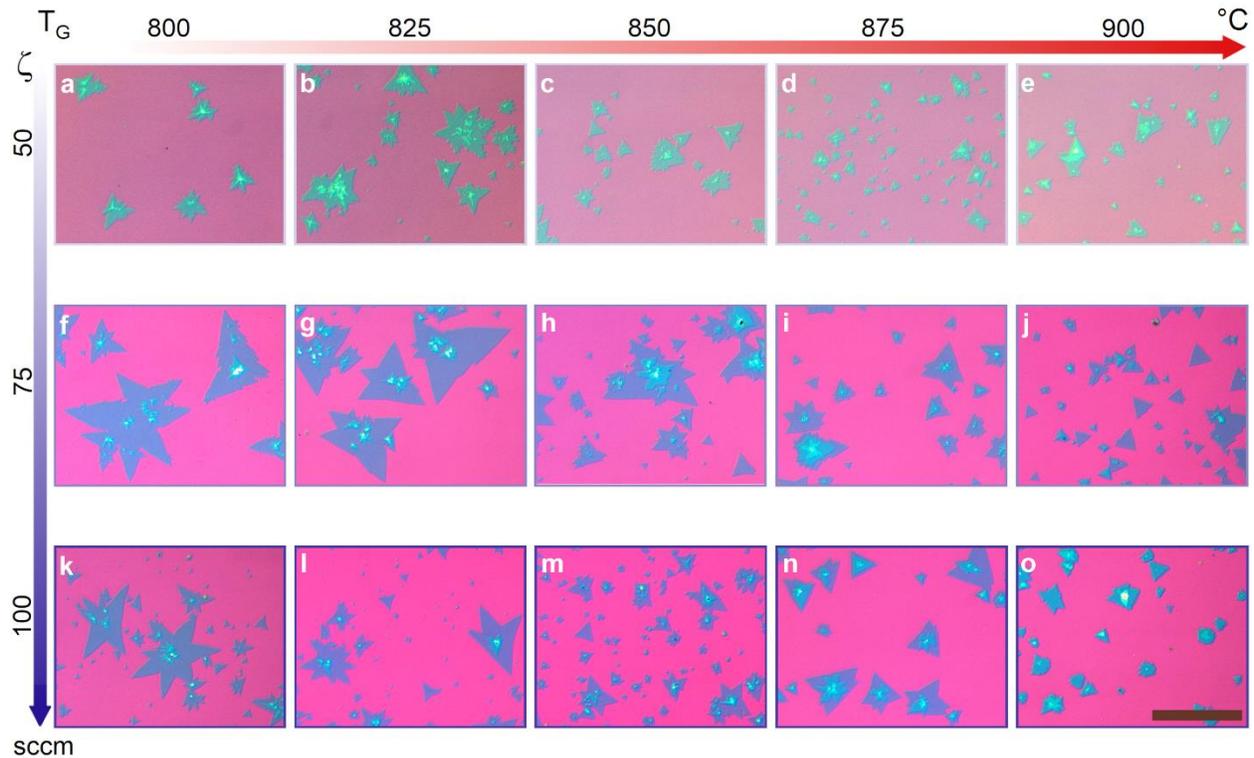

**Fig. 3** OM images of as-grown $MoS_2$ samples as a function of growth temperature ($T_G$) and inert gas (Argon) flow rate ($\zeta$). The sulfur temperature is 140°C. Ideal growth conditions result in OM image (j). Scale bar is 100 μm

Here, instability may occur as atoms do not have enough time to move to the right lattice sites where crystal domains have the lowest surface free energy; hence, the probability of defect formation increases [43]. However, a new balance can be achieved to stabilize the crystal growth and shift the growing conditions to the thermodynamic control by increasing $T_G$. This results in regular shape flake growth, see Fig. 3 (j, n, o). In such conditions, increased evaporation rate of $MoO_3/MoO_x$ within the region of synthesis due to increased $T_G$, followed by enhanced sulfur inflow, leads to morphology restoration. Thus, by controlling the flow rate of the carrier gas during the growth process we can precisely control the reaction process, avoiding blocking effects on $MoO_3/MoO_x$ precursors and control the growth procedure influencing the overall morphology evolution of the samples, directing it toward the desired substrate coverage densities, crystal shapes and lateral dimensions [48]. For our CVD setup, the sample with the best morphology is grown under the following conditions: $T_G= 900°C$, $T_S= 140°C$ and $\zeta= 75$ sccm, see Fig. 3 (j). At this carrier gas flow rate, the density of $MoS_2$ crystals is moderate and the shape of crystals become triangular, suggesting Mo-zigzag (Mo-zz) triangle terminations of flakes [34][43]. These growth parameters give not only the best growth morphology of the sample but, as it will be shown later, it gives the best optical properties of the CVD synthesized monolayer $MoS_2$ flakes.

Figure 4 presents SEM images from the samples synthesized under different growth conditions, proving how the variation of $MoS_2$ shape is closely correlated with the variation in the growth environment. It can be seen that for low inert gas flow rate, high $T_S$ and low $T_G$ values promote dendritic growth due to the high sulfur concentrations, resulting in the irregular crystal shape and substantial lateral sizes, see Fig. 4(a). Not only that, low $T_G$ also promotes multiple growth center activation. Further decreasing of the sulfur solid precursor temperature decreases the sulfur evaporation rate and hence its partial vapor pressure in the chamber, that shifts the geometry from dendritic to the concave triangle, Fig. 4 (b). Rising the growth temperature decreases the concavity of the crystal shape toward the triangles and truncated triangles and the general trend in the evolution of $MoS_2$ morphology with growth conditions can be easily identified: $MoS_2$ increases its compactness and decreases its concavity as the growth temperature and/or Mo/S precursor ratio increase [37].

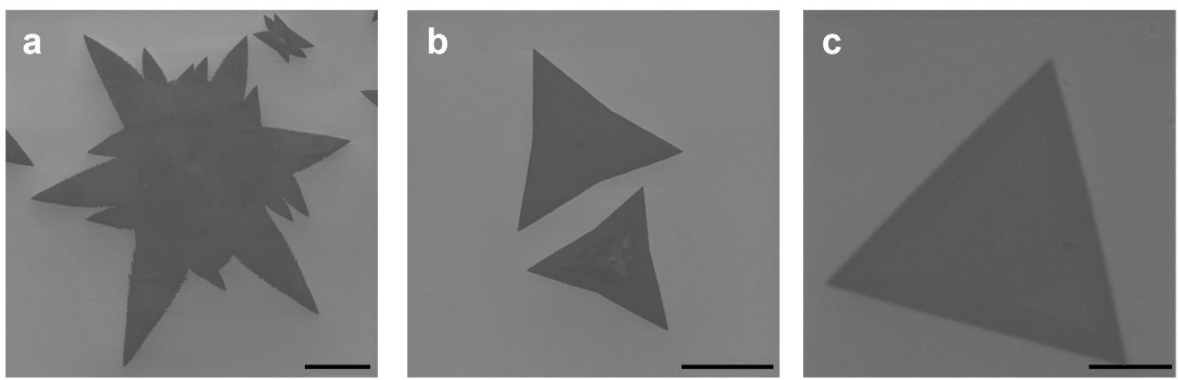

**Fig. 4** SEM images of as-grown $MoS_2$ samples as a function of ($T_G$, $T_S$) for inert gas flow rate $\zeta$=50 sccm. (a) (775°C, 145°C); (b)(775°C,140°C) and (c) (900°C,140°C). Scale bar is: 20 μm for (a) and (b) and 5 μm for (c)

The surface morphology of the grown $MoS_2$ islands was also investigated with AFM, see Fig. 5. Inset shows corresponding height profile along the solid black line. The height difference between the $MoS_2$ island and the substrate is ~ 0.8 nm, which is in agreement with the previously reported monolayer thickness of CVD grown $MoS_2$ [49]. The AFM topography image shows the presence unreacted precursors residues and craters on the wafer surface [50], which are indicated by yellow and black arrows, respectively.  Unreacted precursors appear as white dots with higher topography, increasing the substrate surface roughness (~ 4.9 nm) that is significantly higher than the MoS$_2$ island roughness (~ 0.15 nm). Craters are typically observed in CVD grown MoS$_2$ samples using solution-based precursors with a sodium promoter. Residuals NaOH and NH$_3$ from the chemical reaction described in Eq. 1 may cause etching of SiO$_2$ substrate, and thus produce typically observed craters that are several nm deep [51]. These features can be removed either by a post-synthesis transfer process or by pre-synthesis substrate treatment [50].

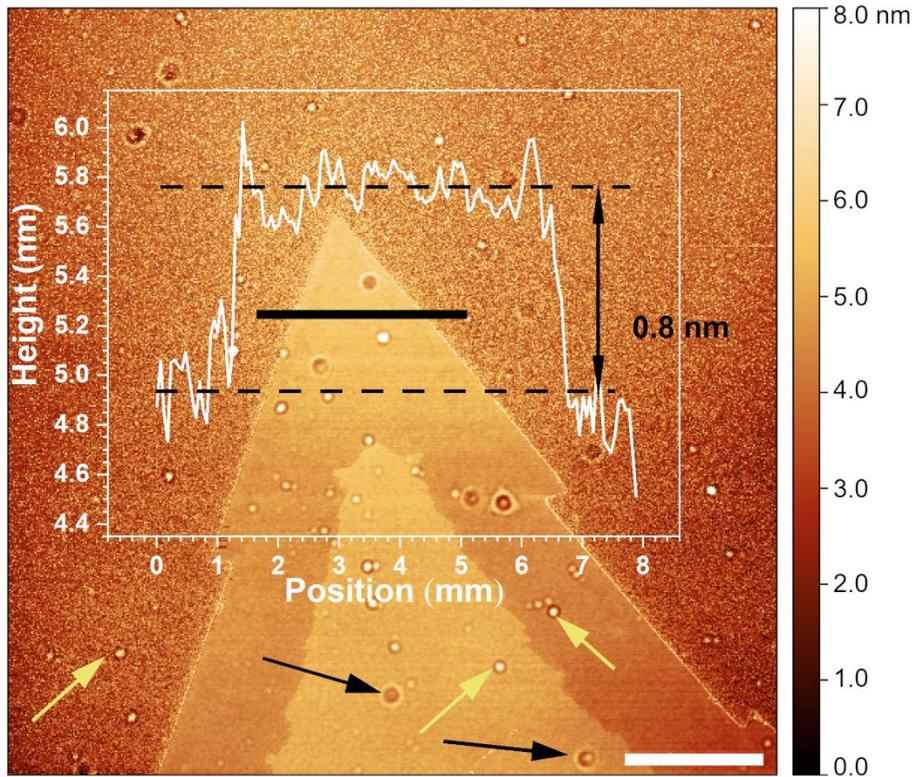

**Fig. 5** AFM topography image of the isolated as-grown MoS$_2$ bilayer obtained for the synthesis parameters: ($T_G$, $T_S$, $\zeta$) = (900°C, 140°C, 75 sccm). Inset shows corresponding height profile along the solid black line. The unreacted precursors and craters on the wafer surface are indicated by yellow and black arrows, respectively

## 2.2. Global optical response

To gain information on how the synthesis parameters influence the global optical response of the grown samples, the PL and Raman spectroscopy point measurements were performed on a randomly chosen flake grown on a specific substrate zone (1-4), for each of 30 samples that were synthesized under certain conditions (see Table S-2 in ESM) to have good unbiased statistics. Systematic data acquisition of the global optical response of samples was done by measuring PL and Raman spectra with respect to the growth parameters for different sample batches (Batch 1- Batch 5) and for different substrate zones (zone 1 – zone 4). Spectra analysis consisted of an automatic fitting procedure that results in a list of fit parameters for the fitting function that gives the best fit of the spectrum. Both the PL and Raman spectra were fitted to two Lorentzian functions. The retrieved fitting parameters of our interest were peak positions, full width at half maximum (FWHM) and the spectral line intensity.

The flake-to-flake differences in the PL intensity response correlates with the exciton lifetime ($\tau_E$) following the relation $I(PL){\sim}\tau_E/\tau_R$ [52], [53] where $\tau_R$ is the radiative recombination time and it is related to exciton lifetime through the relation $\tau_E^{-1} = \tau_R^{-1} + \tau_{NR}^{-1}$. The radiative recombination time is temperature dependent and intrinsic to a specific material, while $\tau_{NR}$ is non-radiative recombination time that depends on a variety of factors such as defect states and traps that shorten non-radiative recombination time, providing more non-radiative channels [54]–[56]. Thus, the PL intensity of both excitonic peaks (A- and B-peaks) is sensitive to changes in the density of defects mediating non-radiative recombination, giving indirect information on sample quality as a result of certain growth conditions.

The information on A exciton intensity was taken as a figure of merit of the sample quality. PL measurements of the grown sample were conducted on the flakes of lateral sizes up to 50 μm. Point PL and Raman measurements were performed close to the flake center, avoiding the nucleation centers. In Fig. 6, the PL spectra of the A exciton as a function of different growth conditions for different batches and different growth temperatures for each of the four substrate zones are presented.

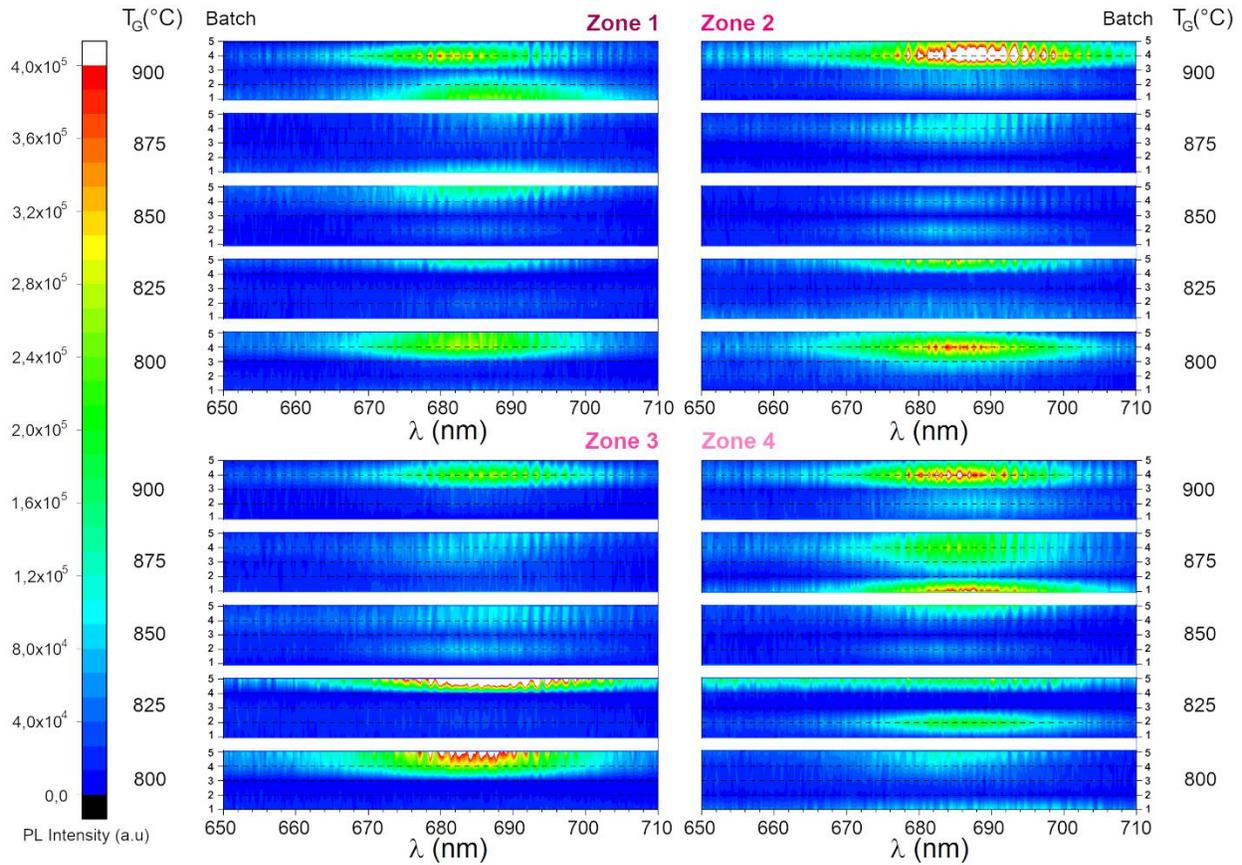

**Fig. 6** The PL spectra of A exciton as a function of different synthesis parameters for all four zones on a substrate

As can be seen from Fig. 6, Batch 4 samples grown under moderate sulfur exposure ($T_S = 140°C$ and $\zeta = 75$ sccm) exhibit strong PL emission peaked at ~ 685 nm. In contrast, samples from Batch 1 grown under the insufficient sulfur exposure ($T_S = 135°C$ and $\zeta = 50$ sccm) exhibit modest PL intensity. Besides the sulfur concentration, the growth temperature significantly influences the overall optical response since a good balance of both precursors in reaction zones plays a crucial role in sample uniformity and quality. A combination of higher $T_G$ and moderate $T_S$ and $\zeta$ results in increased PL emission. Moderate $T_G$ such as 825 and 850°C on average give less intense optical signal for any combination of $T_S$ and $\zeta$. Regarding spatial sample uniformity over the different sample zones, the samples grown in Batch 4 and 5 and for higher $T_G$ give higher optical signals in all four substrate zones. The A exciton peak emission wavelength and full width at half-maximum (FWHM) of the emission peaks are comparable between different sample batches and different zones, see Fig. S-4 (a) and (b), respectively. The average wavelength of the A exciton peak averaged over all fitted values is $(687 \pm 8)$ nm and its averaged FWHM value is $(14 \pm 6)$ nm and comparable with the values from the literature [51], [57], [58]. From Fig. S-4 (a) it can be noticed that in Zone 3 and 4 A exciton peak positions are less dispersed for different $T_G$. Optically homogeneous sample growth in these zones is probably the result of a more convenient position of the contact zone with respect to the sulfur boat, where the inflow of sulfur and reaction between two precursors is well-balanced and results in more steady growth. This is also confirmed with the global FWHM data, see Fig. S-4 (b). It can be generally noticed that samples grown at higher $T_G$ experience less deviation from the mean values for A exciton peak position and spectral line FWHM. This implies that if the favorable growth conditions are established, sulfur vapor concentration gradient along the Si substrates due to its relative distance from the sulfur boat is not crucial for establishing uniform, high crystallinity and high-quality $MoS_2$ growth along the entire substrate.

Many factors may influence the PL signal characteristics such as defects (vacancies, grain boundaries), strain and electrostatic doping [34]. To rule out strain and electrostatic doping as primary causes of the variations in emission intensity, we have simultaneously measured Raman spectra since the frequency of $E^1_{2g}$ vibrational mode is sensitive to strain [59] [60], whereas the frequency of $A_{1g}$ vibrational mode is sensitive to electrostatic doping [60] [61]. The energy positions of Raman vibrational modes for different $T_G$ are presented in Fig. 7 (note that other growth parameters are also different since systematization here is done only by $T_G$ parameter value). $T_G$ increase may induce the strain in $MoS_2$ during the growth on $SiO_2/Si$ wafer during fast-cooling from the growth temperature to 650°C, since $SiO_2$ substrate and $MoS_2$ have different thermal coefficients [62][63].

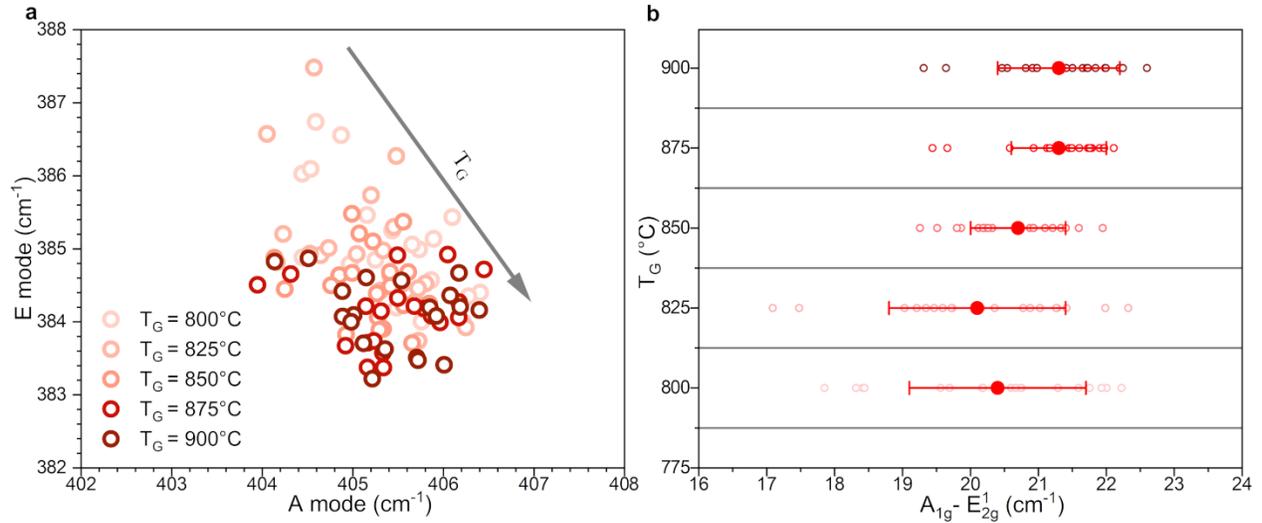

**Fig. 7** The Raman peak position (a) and for the Raman mode peak position difference (b) for different $T_G$

Although the slight softening of the $E^1_{2g}$ mode is observed as $T_G$ increases, we can conclude that strain due to the thermal coefficients mismatch is not playing a crucial role in the overall sample optical response when a proper choice of growth parameters is ensured, see Fig. S-5. It can be noted that the redshift of the $E^1_{2g}$ mode is more pronounced for batch 2 than for batch 4 samples implying that obtained $E^1_{2g}$ mode shift is inherent to CVD grown sample but also it depends on growth conditions. Besides, intensity of the Raman peaks increases for batch 4 samples, indicating better crystal quality [64]. Since the average values of $A_{1g}$ mode energies show no significant change with the growth parameters, the change in the Raman mode peak positions difference with $T_G$ of about 0.9 cm$^{-1}$ (see Fig. 7 b) indicates that both strain and doping can be excluded as the cause of the non-uniform PL intensity between different sample batches [65]. Instead, this behavior points to structural defects as the source of non-uniformity. In general, measured Raman spectra for different growth conditions do not have signatures from oxide phases ($MoO_3$ and/or $MoO_x$) which would indicate the presence of residual oxygen associated with incomplete sulfurization [66]. In the cited paper, X-ray photoelectron spectroscopy shows that an increase in the degree of sulfurization leads to an improvement in the relative stoichiometry and a decrease in the amount of $MoO_3/MoO_x$ within the region probed. Our results imply that sufficient sulfur inflow influences the optical response of the grown crystals where the substantial overall optical response was gained from the samples with higher growth temperatures and intermediate sulfur temperature and intermediate flow. We concluded that for the synthesis parameters: $T_G = 900°C$, $T_S = 140°C$ and $\zeta = 75$ sccm, the sample optical response is homogeneous throughout the substrate. Some other combinations of low growth temperature and medium or high argon flow also yield high optical response, while synthesis with higher $T_G$ ($T_S= 135°C$, $\zeta= 50$ sccm, $T_G= 900°C$ - first batch) and ($T_S= 140°C$, $\zeta= 100$ sccm, $T_G= 900°C$ - fifth batch), generally yields low PL intensity. It implies that increased $T_G$ itself is not sufficient for achieving preferred optical response but also other growth parameters have to be optimized. For the growth temperature $T_G= 900°C$, the peak energies of A and B excitons are nearly constant for all other synthesis parameters ($T_S$ and $\zeta$) throughout the sample (i.e. in all 4 zones the positions of A and B excitons do not change significantly, see Fig. S-6.)

## 2.3. Local optical response

To obtain information on the local optical properties of individual $MoS_2$ islands and their relation to growth parameters, room temperature PL optical maps of samples from the Batch 4 were taken. As shown in the previous section, samples grown under $T_S = 140°C$ and $\zeta = 75$ sccm flow growth conditions exhibit strong PL emission. Sample uniformity is investigated across representative

~15-50 μm triangular flakes, grown at different $T_G$, i.e. for different stoichiometric ratio. The optical PL maps presented in Fig. 8 show the PL intensity and peak central wavelengths for MoS$_2$ grown for $T_G$ at 800°C (a and b) and 900°C (d and e), respectively.

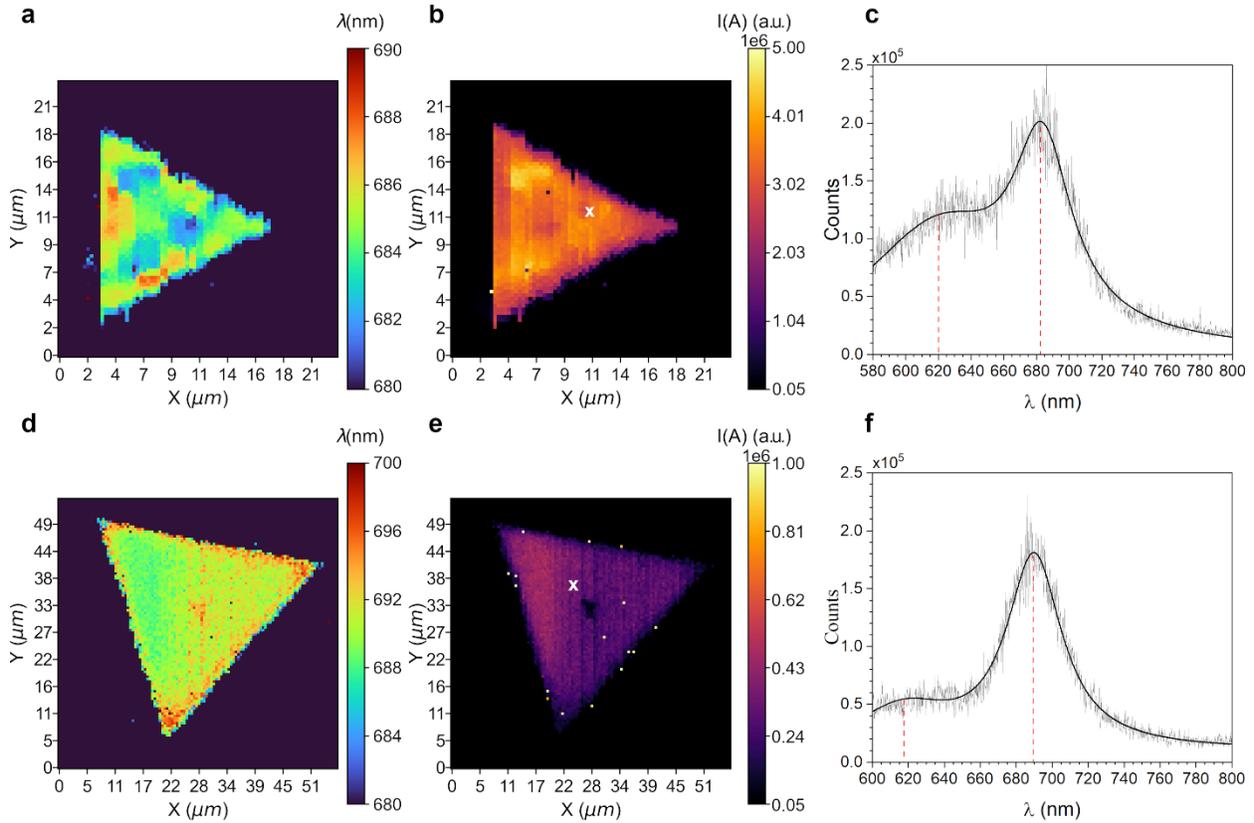

**Fig. 8** Spatial distribution of A exciton peak position (a, d) and intensity (b, e) for samples grown at $T_G = 800°C$ (top row) and $T_G = 900°C$ (bottom row), $T_S = 140°C$, $\zeta = 75$ sccm. Point photoluminescence spectra (c, f) are taken from the inner part of the flake, indicated by white X marks

The PL intensity map for the sample grown at $T_G=800°C$ (Fig. 8 (b)) exhibits variation across the sample. The perimeter, vertices and even some central parts of the flake exhibit lower PL intensity, with high intensity regions that extend radially outward toward the three corners. The variation in PL intensity is accompanied with variation in the peak position (Fig. 8 (a)). We observe that with the increase in growth temperature, spectral maps become more homogeneous but also redshifted, see Fig. 8 (d) and (e). The intensity map for the flake grown at $T_G$=900°C shows variation in signal intensity due to the defocusing that occurred during the optical mapping of the flake. In Fig. S-7 in SI optical maps for the other samples (i.e. other $T_G$) from the batch 4 are presented. Even though all these samples have favorable equilateral triangle morphology, we observe from the optical response that a higher growth temperature yields a more homogeneous optical response over the entire sample. The increased evaporation rate of $MoO_3/MoO_x$ within the region of synthesis, due to increased $T_G$ accompanied by adequate sulfur inflow, leads to desired sample quality [64]. Thus, by controlling the flow rate of the carrier gas and growth temperature during the growth process we can precisely control the reaction process, avoiding blocking effects on $MoO_3/MoO_x$ precursors. At the same time, control of the growth procedure influences the overall morphology evolution of the samples, directing it toward not only favorable morphology in the sense of crystal shapes and lateral dimensions [50] but also towards uniform optical response as an indication of high-quality electronic properties. Again, for the synthesis parameters: $T_G$ = 900°C, $T_S$ = 140°C and $\zeta$ = 75 sccm, we concluded that the sample has homogeneous local optical response throughout the whole flake [64].

In order to investigated the $MoS_2$-substrate interaction and the influence of interfacial effects on PL homogeneity of this particular sample as well as adsorption and desorption mechanisms, we tested our sample on the presence of trions. It is known from the literature that a high intensity laser can enhance trion as well as induce biexciton emission in TMD [58], [67]–[69]. An investigation of the relationship between emission spectra and laser excitation power is presented in Fig. 9. As is evident from the normalized PL spectra of $MoS_2$ emission (Fig. 9(a)), there are two peaks for all incident powers spanning over three orders of magnitude, consistent with emission from the neutral excitons A and B centered at (697.3 ± 0.8) nm and (633 ± 2) nm, respectively, see

Fig. 9(b). The integrated PL intensity is obtained within the range from 575 nm to 815 nm depends linearly on the power (Fig. 9(c)).

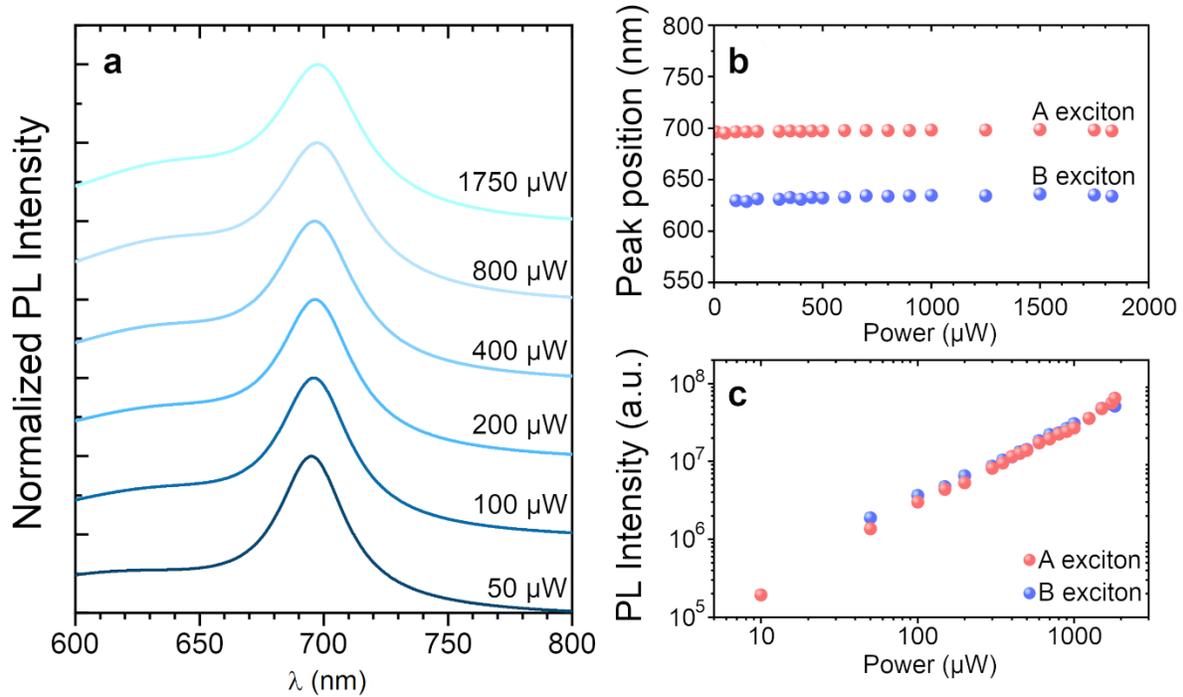

**Fig. 9** Photoluminescence laser power dependence in ambient conditions for MoS$_2$ island grown at (T$_G$, T$_S$, ζ) = (900°C, 140°C, 75 sccm). Both the PL spectral lineshapes (a) and peak positions (b) are unaffected by increased laser power. PL intensity show linear increase with laser power increase (c)

Since the MoS$_2$ sample grown at (T$_G$, T$_S$, ζ) = (900°C, 140°C, 75 sccm) shows no variation in spectral shape or position of PL spectra with an increase in the laser power we assume that the observed behavior is due to the interfacial effects between MoS$_2$ and supporting substrates. Better MoS$_2$ - substrate contact promotes efficient electron transfer from the MoS$_2$ to the substrate, hindering the trion formation [70] and adsorption and desorption mechanisms of O$_2$ containing species. Moreover, the increase of the laser power during the power-dependent measurements does not influence PL spectral lineshape, indicating no sign of local heating effect [64]. The linearity of emission from the neutral exciton in power dependence indicates the absence of exciton-exciton recombination processes.

## 3. Conclusion

In conclusion, we have studied the effects of the CVD growth parameters on the monolayer $MoS_2$ flakes morphology and their global and local optical properties. We developed a wafer-scale controlled synthesis of $MoS_2$ monolayers, which utilizes a liquid based molybdenum precursor. Based on the variation in the growth temperature, that influences Mo-precursor concentration, sulfur temperature and carrier gas flow, which influence sulfur concentration, we have optimized the growth parameters to obtain regular crystalline growth of the samples with a desired optical response not only over the single, randomly chosen flake but also over the whole substrate. The microstructure and optical properties of the $MoS_2$ flakes were characterized by OM, SEM, AFM, Raman and PL spectroscopy techniques for a broad range of the growth parameters. Such a data-driven approach that correlates growth conditions with crystal morphology and its optical response provides us with a practical and necessary pathway to address the challenges towards the controlled synthesis of 2D TMDs and their alloys.

In order to further explore the influence of the growth parameters on the optical properties of CVD grown samples characterized here, we have extended our research toward low-temperature optical measurements, where defect-bound excitons provide information on defect states occurrence. Although detailed discussion is beyond the scope of this paper, our results confirm not only high-quality of grown materials but also its long-term stability even after extensive exposure to vacuum conditions and broad temperature range. Hence, our work adds value to the existing knowledge on the reproducible synthesis of high-quality $MoS_2$ monolayers, which have potential to provide insights into fundamental scientific research and technological applications.

## Acknowledgements

This work was supported by the Croatian Science Foundation, Grant No. UIP-2017-05-3869 and Center of Excellence for Advanced Materials and Sensing Devices, ERDF Grant No. KK.01.1.1.01.0001. Authors gratefully acknowledge support from dr. Nikša Krstulović (Institute of Physics, Zagreb) for providing access to the optical microscope (Croatian Science Foundation, Grant No. PZS-2019-02-5276).

**Authors Contributions:** Ana Senkić (orcid.org/0000-0002-0567-5299): Conceptualization, Methodology, Investigation- synthesis, PL and Raman measurements, Systematization, Visualization, Writing - Original Draft; Josip Bajo (orcid.org/0000-0002-8381-5488): Conceptualization, Methodology, Investigation- synthesis, PL and Raman measurements, Systematization, Visualization, Writing - Original Draft; Antonio Supina (orcid.org/0000-0002-3990-5830): Investigation- synthesis and AFM measurements, Writing - Review & Editing**;** Borna Radatović (orcid.org/0000-0001-5012-6005): Investigation- SEM measurements, Writing - Review & Editing**;** Nataša Vujičić (orcid.org/0000-0002-5437-5786): Supervision, Funding acquisition, Conceptualization, Systematization, Visualization, Writing - Original Draft.

Ana Senkić and Josip Bajo contributed equally to this work.

## Declaration of Competing Interest

The authors declare no competing financial interest.

## Data availability

Data will be made available on request.

**Supplementary Information Available:** Supplementary Information includes: (i) description of the growth density of the samples (ii) growth conditions for all the samples demonstrated in the manuscript, (iii) flake morphology and growth density for broad $T_S$ span; (iv) PL intensity and FWHM of all analyzed flakes (different batches, different growth condition and different zones); (v) Raman spectra for two different sample batches and different $T_G$; (vi) A and B exciton peak energies for $T_G$= 900°C for all batches and (vii) the increase in the local sample uniformity with $T_G$ for the batch 4 samples. This material is available in *Electronic Supplementary Material.pdf* file.

# Electronic Supplementary Material

## Effects of CVD Growth Parameters on Global and Local Optical Properties of MoS₂ Monolayers

Ana Senkić[1,2*·], Josip Bajo[3·], Antonio Supina[1,2], Borna Radatović[1], and Nataša Vujičić[1*]

[1] *Institute of Physics, Center for Advanced Laser Techniques and Center of Excellence for Advanced Materials and Sensing Devices, Zagreb 10 000, Croatia*
[2] *Faculty of Physics, University of Rijeka, Rijeka 51 000, Croatia*
[3] *University of Vienna, Faculty of Physics, Vienna Center for Quantum Science and Technology (VCQ), 1090 Vienna, Austria*

*Address correspondence to Ana Senkić, asenkic@ifs.hr; Nataša Vujičić, natasav@ifs.hr
· These authors contributed equally to this work.

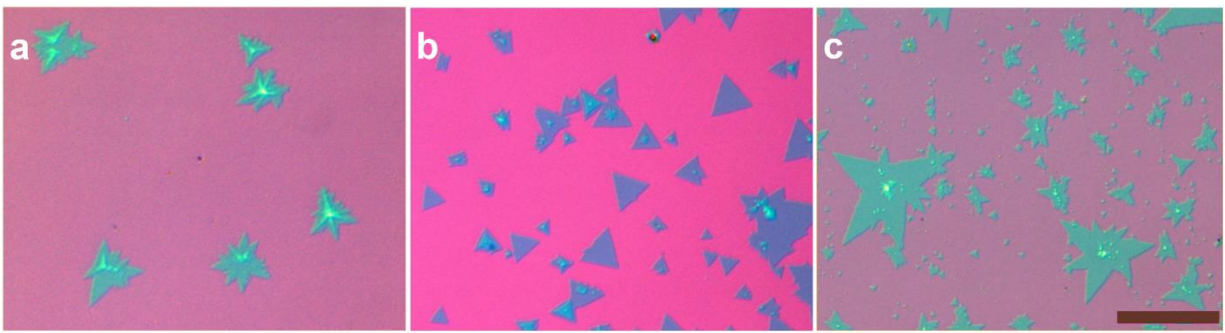

**Fig. S-1** Poor (a), moderate (b) and high (c) sample coverage due to the influence of synthesis parameters on the lateral growth. Scale bar is 100 μm.

| BATCH No. | Ts (°C) | ζ (sccm) | T_G (°C) |
|---|---|---|---|
| 1 | **135** | 50 | 775 |
| | | | 800 |
| | | | 825 |
| | | | 850 |
| | | | 875 |
| | | | 900 |
| 2 | **140** | 50 | 775 |
| | | | 800 |
| | | | 825 |
| | | | 850 |
| | | | 875 |
| | | | 900 |
| | | | 925 |
| 3 | **145** | 50 | 775 |
| | | | 800 |
| | | | 825 |
| | | | 850 |
| | | | 875 |
| | | | 900 |
| | | | 925 |
| 4 | **140** | 75 | 800 |
| | | | 825 |
| | | | 850 |
| | | | 875 |
| | | | 900 |
| 5 | **140** | 100 | 800 |
| | | | 825 |
| | | | 850 |
| | | | 875 |
| | | | 900 |

**Table S-2:** Growth parameters for 30 samples sorted into 5 batches

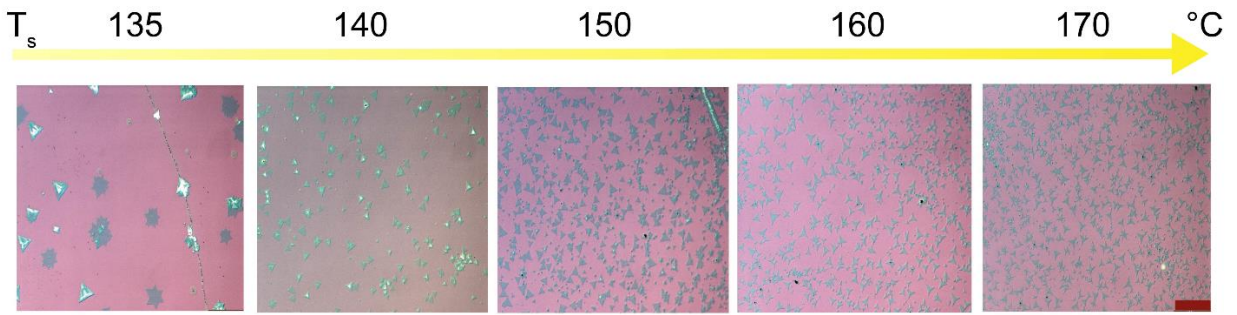

**Fig. S-3** Optical micrographs of the samples grown for a broader $T_S$ range (from 135 up to 170°C). $T_S$ increase causes dense growth of the three-point star shape MoS$_2$ flakes. Adopted from [40]

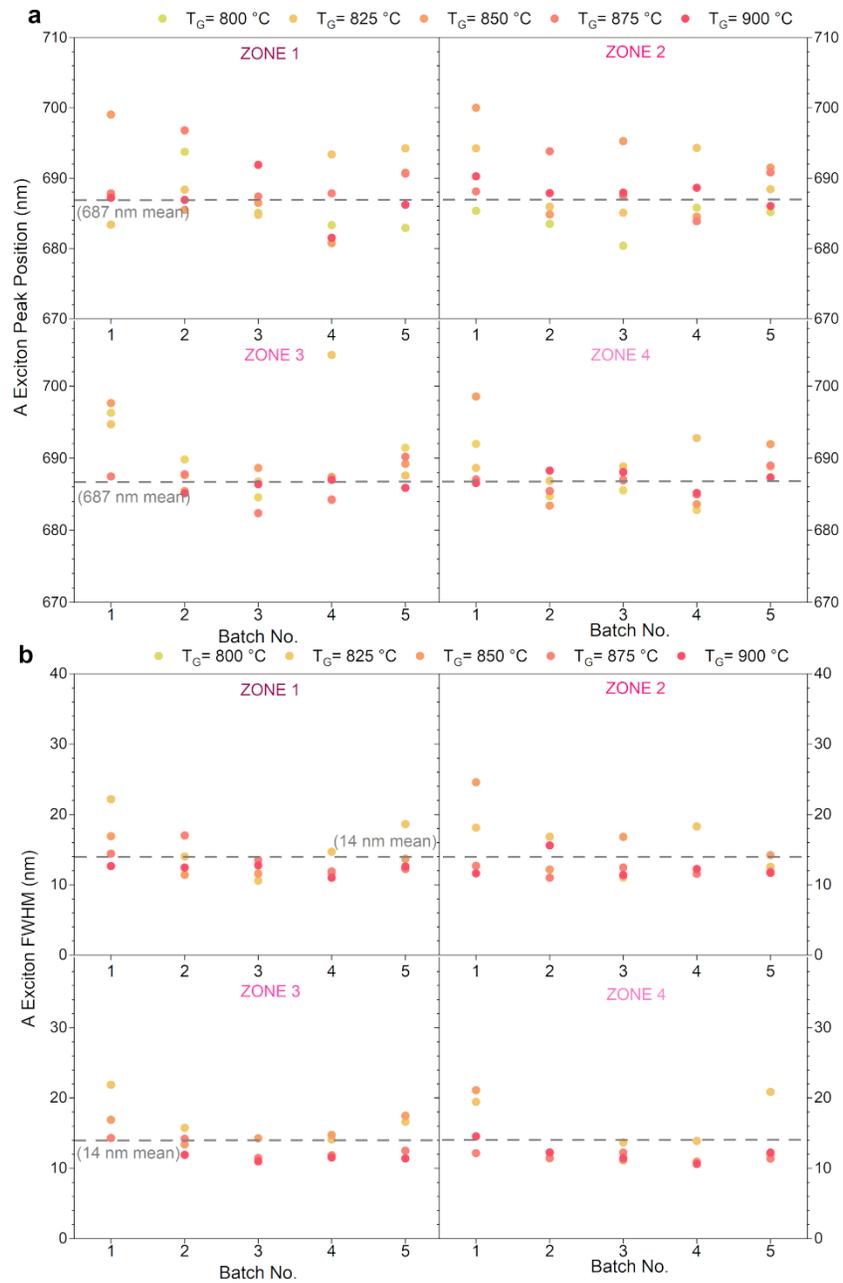

**Fig. S-4**: (a) A exciton peak position and (b) FWHM of exciton A emission lines for different batches and different wafer reaction zones.

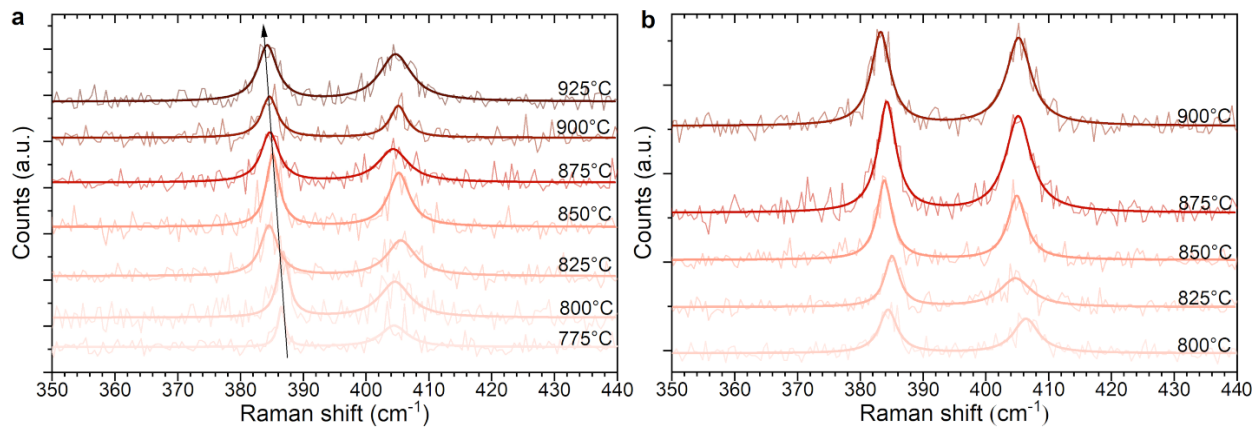

**Figure S-5**: Raman spectra for different $T_G$ for Batch 2 (a) and Batch 4 (b) samples. The shift of E Raman mode with the $T_G$.

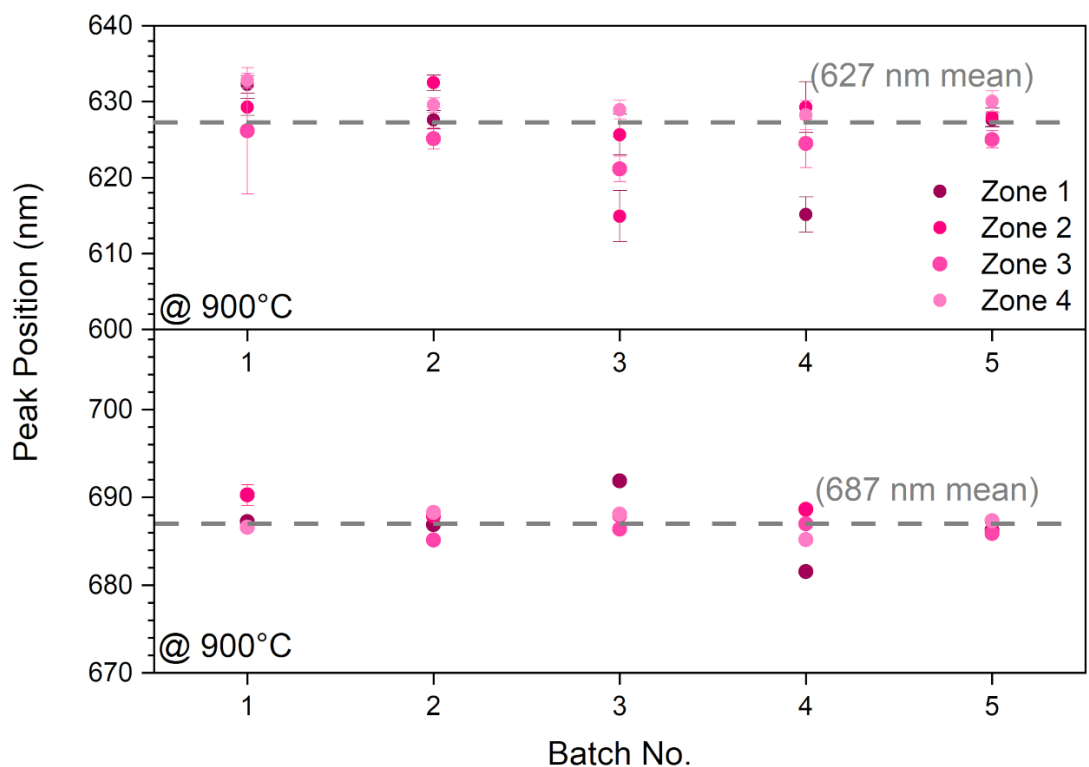

**Figure S-6**: A and B exciton peak energies for $T_G$= 900°C. The peak energies are nearly constant for all other synthesis parameters ($T_S$ and $\zeta$) throughout the sample (i.e. in all 4 zones)

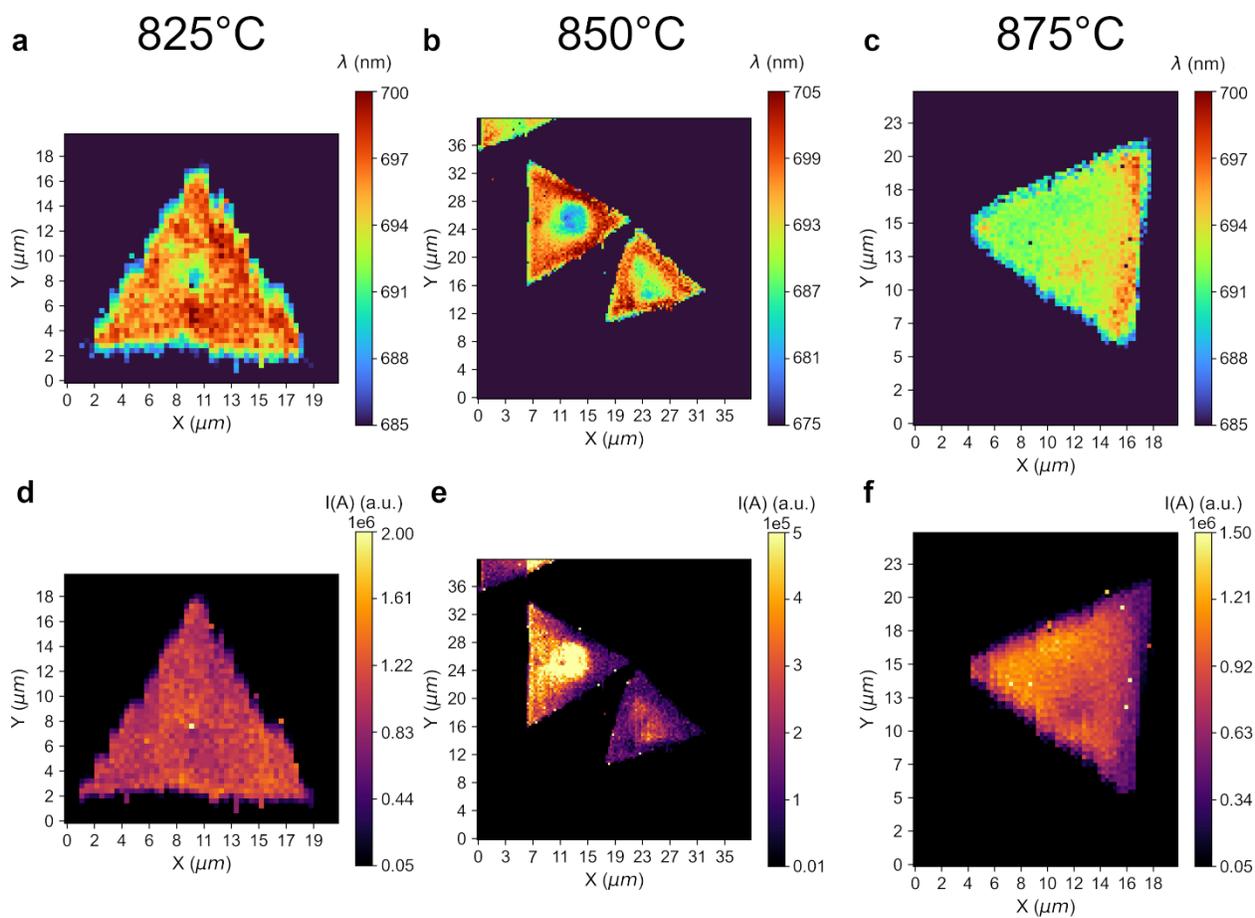

**Figure S-7**: The increase in the local sample uniformity with $T_G$ both in spectral (a-c) and intensity (d-e) response. Presented flakes are from the batch 4.